\documentclass[fleqn,usenatbib]{mnras}

\usepackage{newtxtext,newtxmath}

\usepackage[T1]{fontenc}
\usepackage{ae,aecompl}

\usepackage{graphicx}	
\usepackage{amsmath}	
\usepackage{amssymb}	

\usepackage[usenames,dvipsnames]{xcolor}

\newcommand{\bz}{$\langle B_z \rangle$}	
\newcommand{\logg}{log $g$}

\newcommand{\kms}{$\text{km } \text{s}^{-1}$}

\title[A sudden change of the magnetic field of AD Leo]{A sudden change of the global magnetic field of the active M dwarf AD Leo revealed by full Stokes spectropolarimetric observations\thanks{Based on observations obtained at the Canada-France-Hawaii Telescope (CFHT) which is operated by the National Research Council of Canada, the Institut National des Sciences de l'Univers of the Centre National de la Recherche Scientifique of France, and the University of Hawaii.}}

\author[A. Lavail et al.]{
A.~Lavail$^{1}$\thanks{E-mail: alexis.lavail@physics.uu.se},
O.~Kochukhov$^{1}$,
G.~A.~Wade$^{2}$
\\
$^{1}$Department of Physics and Astronomy, Uppsala University, Box 516, 751 20 Uppsala, Sweden\\
$^{2}$Department of Physics and Space Science, Royal Military College of Canada, PO Box 17000, Stn Forces, Kingston, \\ Ontario K7K 7B4, Canada
}

\date{Accepted 2018 July 6. Received 2018 June 11; in original form 2018 April 6}

\pubyear{2018}

\begin{document}
\label{firstpage}
\pagerange{\pageref{firstpage}--\pageref{lastpage}}
\maketitle

\begin{abstract}
In this paper we present an analysis of the first high-resolution full Stokes vector spectropolarimetric observations of the active M dwarf AD~Leo. Based on observations collected in 2016 with the ESPaDOnS instrument at CFHT, we derived the least-squares deconvolved Stokes profiles and detected linear polarisation signatures in spectral lines. At the same time, we discovered that the circular polarisation profiles corresponding to our data set are significantly weaker compared to all archival spectra of AD~Leo, which exhibited approximately constant profiles over the timescale of at least 6 years until 2012. Magnetic maps obtained using Zeeman Doppler imaging confirm the sudden change in the surface magnetic field. Although the total magnetic field energy decreased by about 20\% between 2012 and 2016, the field component responsible for the observed circular polarisation signatures corresponds to a stronger field occupying a smaller fraction of the stellar surface in the more recent map. These results represent the first evidence that active M dwarfs with dipole-dominated axisymmetric field topologies can undergo a long-term global magnetic variation.
\end{abstract}




\section{Introduction}
\label{section:introduction}
Magnetic fields play a key role at all stages of stellar evolution. While magnetic fields are believed to influence and even rule physical processes inside and around stars -- such as convection, accretion, rotation rates, flares, and winds among others -- the origin and mechanism behind the generation of magnetic fields is not thoroughly understood. In the Sun and solar-like stars, it has been proposed that the magnetic field is generated and maintained by a so-called $\alpha \Omega$ dynamo \citep{1955ApJ...122..293P}, representing an interplay of differential rotation and turbulence due to convection around the tachocline region. Fully convective stars do not harbour a tachocline but they have been detected to host strong magnetic fields \citep{2006Sci...311..633D,2008MNRAS.390..567M,2017NatAs...1E.184S}, that can be reasonably reproduced with ``$\alpha^2$ dynamo'' -- solely driven by turbulence \citep{2015ApJ...813L..31Y}.

AD Leo (GJ 388, BD +20 2465) is an extensively studied active M3Ve dwarf at the threshold of the fully convective regime. It emits strong \citep{1991ApJ...378..725H} and frequent flares \citep[][and references therein]{2006A&A...452..987C,2012PASP..124..545H}.

The magnetic field of AD Leo has been repeatedly studied through the analysis of intensity \citep{1996ApJ...459L..95J,2000ASPC..198..371J,2007ApJ...656.1121R,2010A&A...523A..37S} and circularly polarised spectra \citep{2008MNRAS.390..567M}. The latter study suggests that the large-scale magnetic field of AD~Leo is relatively simple -- predominantly a dipole aligned with the rotation axis and seen close to the pole -- and with a very limited intrinsic variability during the period spanned by their observations (2007--2008).

We present here an analysis of spectropolarimetric observations of AD~Leo in all four Stokes parameters that resulted in the detection of linear polarisation signatures in spectral lines, and the discovery of a significant change in the shape of the circular polarisation profiles, which we attribute to an important change in the surface magnetic field of the star. We describe our observations in Sect.~\ref{section:observations}. The comparison with previous observations and surface magnetic field mapping is presented in Sect.~\ref{section:analysis}, and we finally discuss our results in Sect.~\ref{section:discussions}.


\begin{table*}
	\centering
	\caption{Log of the AD Leo spectropolarimetric observations. Columns 1--6 list, respectively, the UTC date of observation, the heliocentric Julian date, the Stokes parameter observed, the exposure time, the signal-to-noise ratio of polarised spectra, and the signal-to-noise ratio of LSD Stokes profiles. Column~7 gives the rotation cycle starting from the first observation,  column 8~catalogues mean longitudinal magnetic field {\bz} values, column 9~lists the LSD False Alarm Probability (FAP), and column~10 indicates if the detection of polarisation signature in the LSD profiles is definite (DD), marginal (MD), or if there is no detection (ND).}
	\label{table:observation_log}
	\begin{tabular}{ccccccccll} 
		\hline
		Date		& HJD		&Stokes	& $t_\text{exp}$	& S/N spectra & S/N LSD	& Rotational & \bz & FAP & Detection\\
        (UTC)		& ($2453000 +$)	& &(s)			&		& &	cycle		&(G) 			& 					&\\
		\hline
		2016-02-17	& 4435.7957 & $V$	& $4\times 600$	&386	& $19.0\times 10^3$	& 00.000	& $-182\pm 8$	& $0$ 				&DD\\
		2016-02-17	& 4435.8269 & $Q$ &	 $4\times 600$	&377	& $27.2\times 10^3$	& 00.014	& - 			& $3\times 10^{-3}$	&ND\\
		2016-02-17	& 4435.8580 & $U$ & $4\times 600$	&393	& $28.3\times 10^3$	& 00.028	& - 			& $8\times 10^{-1}$	&ND\\
        2016-02-18	& 4436.8831 & $V$ & $4\times 600$	&380	& $18.6\times 10^3$	& 00.485	& $-176\pm 8$	& $0$				&DD\\
        2016-02-18	& 4436.9141 & $Q$ & $4\times 600$	&432	& $31.3\times 10^3$	& 00.499	& - 			& $4\times 10^{-3}$	&ND\\
        2016-02-18	& 4436.9452 & $U$ & $4\times 600$	&424	& $31.1\times 10^3$	& 00.513	& - 			& $2\times 10^{-4}$	&MD\\
        2016-02-23	& 4441.8954 & $V$ & $4\times 600$	&384	& $18.6\times 10^3$	& 02.723	& $-201\pm 8$	& $0$				&DD\\
        2016-02-23	& 4441.9259 & $Q$ & $4\times 600$	&442	& $31.2\times 10^3$	& 02.737	& -				& $4\times 10^{-6}$	&DD\\
        2016-02-23	& 4441.9562 & $U$ & $4\times 600$	&450	& $31.7\times 10^3$	& 02.750	& -				& $1\times 10^{-15}$&DD\\
        2016-02-24	& 4443.0074 & $V$ & $4\times 600$	&410	& $19.3\times 10^3$	& 03.220	& $-178\pm 7$	& $0$				&DD\\
        2016-02-24	& 4443.0377 & $Q$ & $4\times 600$	&433	& $31.1\times 10^3$	& 03.233	& -				& $6\times 10^{-5}$	&MD\\
        2016-02-24	& 4443.0723 & $U$ & $4\times 600$	&405	& $29.4\times 10^3$	& 03.249	& -				& $1\times 10^{-5}$	&MD\\
		2016-02-29	& 4447.9103 & $V$ & $4\times 300$	&298	& $14.5\times 10^3$	& 05.409	& $-172\pm 10$	& $0$				&DD\\
        2016-03-01	& 4449.0011 & $V$ & $4\times 300$	&266	& $12.8\times 10^3$	& 05.896	& $-200\pm 11$	& $0$				&DD\\
        2016-03-02	& 4449.9154 & $V$ & $4\times 300$	&311	& $14.4\times 10^3$	& 06.304	& $-159\pm 10$	& $0$				&DD\\
        2016-03-03	& 4450.8328 & $V$ & $4\times 300$	&278	& $13.9\times 10^3$	& 06.713	& $-215\pm 10$	& $0$				&DD\\
        2016-04-17	& 4495.8253 & $V$ & $4\times 600$	&226	& $11.3\times 10^3$	& 26.800	& $-211\pm 12$	& $0$				&DD\\
        2016-04-17	& 4495.8564 & $Q$ & $4\times 600$	&366	& $26.9\times 10^3$	& 26.814	& -				& $0$				&DD\\
        2016-04-17	& 4495.9021 & $U$ & $8\times 600$	&289	& $20.4\times 10^3$	& 26.834	& -				& $9\times 10^{-4}$	&MD\\
        2016-04-20	& 4498.7442 & $V$ & $4\times 600$	&343	& $16.9\times 10^3$	& 28.103	& $-188\pm 9$	& $0$				&DD\\
        2016-04-20	& 4498.7754 & $Q$ & $4\times 600$	&411	& $29.7\times 10^3$	& 28.117	& -				& $0$				&DD\\
		2016-04-20	& 4498.8058 & $U$ & $4\times 600$	&438	& $30.7\times 10^3$	& 28.131	& -				& $3\times 10^{-8}$	&DD\\
		\hline
	\end{tabular}
\end{table*}

\section{Observations}
\label{section:observations}

We acquired spectropolarimetric observations of AD~Leo with the ESPaDOnS spectropolarimeter at the 3.6-m Canada-France-Hawaii Telescope. ESPaDOnS is a fibre-fed cross-dispersed spectrograph mounted at the Cassegrain focus of the telescope, with the ability to record circularly and linearly polarised spectra from about 370 to 1000 nm in a single exposure with a resolving power of $R \sim 65000$ \citep{donati2003-espadons,donati++2006-espadons}. The ESPaDOnS polarimeter module consists of a set of Fresnel rhombs -- one fixed quarter-wave rhomb located between two rotating half-wave rhombs -- and of a Wollaston prism splitting the incoming beam into two orthogonal polarisation states. The rotation of the two half-wave rhombs with respect to the other elements allows to switch between the analysis of linear or circular polarisation, but also to swap the location of the two polarised beams inside the instrument, hence mitigating instrumental polarisation signatures. Each Stokes $V$, $Q$, or $U$ observation consists of a set of four consecutive sub-exposures taken with two different polarimeter configurations, allowing for a first-order correction of spurious signals using the ratio method \citep{donati++1997-libre-esprit}. The polarized spectra were reduced at the telescope by the \texttt{UPENA} pipeline\footnote{\url{ http://www.cfht.hawaii.edu/Instruments/Upena/}} running the \texttt{LIBRE-ESPRIT} software \citep{donati++1997-libre-esprit}.

\begin{figure*}
	\includegraphics[width=1.7\columnwidth]{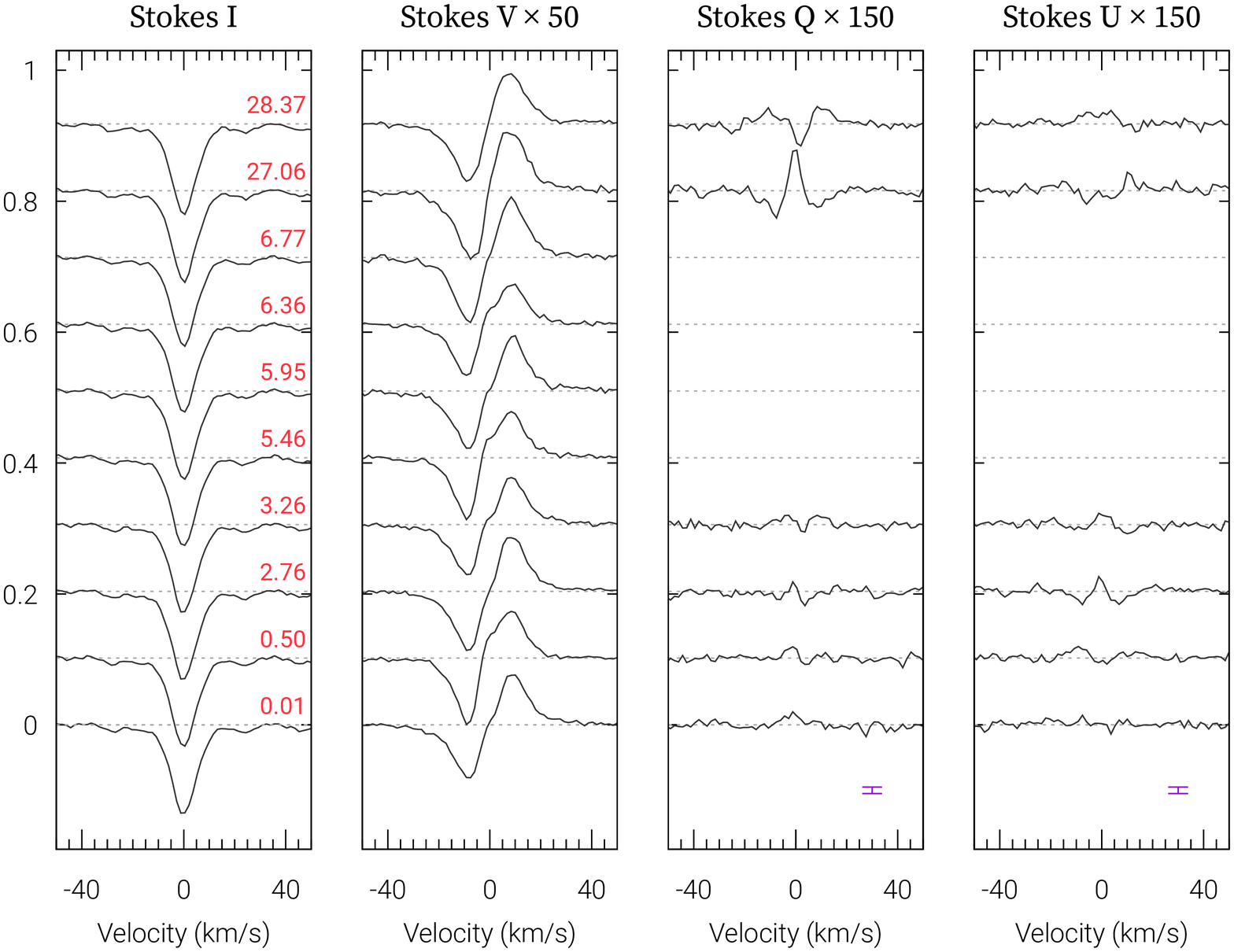}
    \caption{Time series of the full Stokes vector LSD profiles derived from our observations of AD~Leo. The LSD profiles are shifted vertically by a constant offset, according to their rotational cycle increasing upwards. The mean rotational cycle for each observing epoch is indicated in red above the Stokes $I$ LSD profiles, with the reference time being the middle of the first Stokes $V$ exposure. The Stokes $I$ LSD profiles have been renormalised as discussed in the text. The median error bars for the Stokes Q and U LSD profiles are shown in purple at the bottom right of their respective plot.}
    \label{fig:fullstokes}
\end{figure*}

Our observations were obtained between February and April 2016. Data were acquired over three periods. The first data acquired were four full Stokes epochs awarded to our original observing proposal. After the discovery of an unexpected change in the amplitude and shape of the Stokes $V$ profiles compared to archival data, we requested and were awarded four additional epochs limited to circular polarisation observations through Director's Discretionary Time (DDT) immediately after the initial observations. Finally, we obtained two full Stokes epochs one month after the end of the DDT observations through an unplanned ESPaDOnS observing run, while another instrument at the telescope was not operational. In total, we collected 10 Stokes $V$, 6 Stokes $Q$ and 6 Stokes $U$ observations.

The log of our observations is presented in Table \ref{table:observation_log}. The signal-to-noise ratios (SNRs) per spectral pixel of the polarised spectra, measured at 666~nm, typically range between roughly 220 and 450, with a median SNR of 390. The rotational cycle $E$ for each observation was computed using the formula 
\begin{equation}
	E = (\text{HJD} - \text{HJD}_0) / P_{\text{rot}} 
\end{equation} 
where HJD and $\text{HJD}_0$ are the heliocentric Julian dates of the given observation and of the first observation respectively -- as listed in the second column of Table~\ref{table:observation_log}, and $P_{\text{rot}} = 2.2399$ days is the rotation period of AD Leo determined by \citet{2008MNRAS.390..567M}.

We used the PolarBase database \citep{2014PASP..126..469P} to download additional archival Stokes $V$ observations of AD~Leo. In total, we retrieved 7 observations acquired with ESPaDOnS in 2006, and 32 observations with Narval -- its twin instrument mounted on the 2-m T\'elescope Bernard Lyot -- from 2007, 2008, and 2012.

\section{Analysis}
\label{section:analysis}
\subsection{Multi-line methods}
\label{subsection:lsd}
The least-squares deconvolution \citep[LSD,][]{donati++1997-libre-esprit,2010A&A...524A...5K} technique was applied consistently to both archival data and our own observations. LSD effectively combines a subset of photospheric atomic lines into an average profile yielding a substantially higher SNR. We extracted the LSD line list from the {\tt VALD3} database \citep{vald2015}, for the stellar parameters $T_{\text{eff}}= 3300~\text{K}$ and {\logg}$~=~5.0$, using a solar metallicity model atmosphere from the MARCS grid \citep{marcs2008}. The line list contains around 1400 atomic lines, between 450 and 985 nm, selected to be stronger than 20\% of the continuum level. No molecular lines were included in the LSD mask because, with a few exceptions, these lines are very weakly sensitive to the field and generally do not exhibit simple linear Zeeman effect. 

The LSD profiles were calculated adopting a normalisation wavelength ${\lambda}_0 = 670 \text{ nm}$ and an effective Land\'e factor $g_{\text{eff}} = 1.2$, which are close to the weighted average over the corresponding parameters of the lines included in the mask. We have also used a second more restrictive line list, excluding several regions in the spectra which were contaminated by telluric lines, which contains around 1150 spectral lines. The LSD polarisation profiles computed with the first line list have a SNR ranging between 11300 and 31700 (and listed in column 6 of Table~\ref{table:observation_log}), which is 38\% greater on average than for the line list excluding telluric regions. The LSD profiles generated from the first line list were subsequently used for calculation of the false alarm probability (FAP) and detection diagnostics, reported in columns 9 and 10 of Table~\ref{table:observation_log}, as they have the best SNR and improve the chances of polarisation signature detection. However, as they were computed including  wavelength regions that are contaminated by tellurics, no corresponding LSD Stokes $I$ profile can be derived, rendering quantitative interpretation of these polarisation profiles impossible. Therefore, we used the lower SNR LSD profiles generated from the second, more restrictive, line list to compute the mean longitudinal magnetic field (Section~\ref{subsection:bz}) and as input to the Zeeman Doppler Imaging inversions (Section~\ref{subsection:zdi}).

The time series of our full-Stokes LSD profiles is shown in Fig. \ref{fig:fullstokes}. The Stokes $I$ LSD profiles exhibited a pseudo-continuum level below unity -- attributed to the contribution of molecular spectral lines not included in the line list -- and have been scaled to bring the pseudo-continuum level to unity. The Stokes $V$, $Q$, and $U$ LSD profiles have been scaled by the same factor.

False alarm probabilities \citep[FAP,][]{1992A&A...265..682D} have been computed for every LSD polarisation profile. The FAP assessment is based on $\chi^2$ statistics and is commonly used to determine how robustly a signal is detected over the noise level in LSD profiles. Detections are usually considered to be definite for FAP values below $10^{-5}$. The detection is marginal for FAP values between $10^{-5}$ and $10^{-3}$. Finally, we consider that there is no detection when FAP values are above $10^{-3}$. 

\begin{figure*}
	\includegraphics[width=1.9\columnwidth]{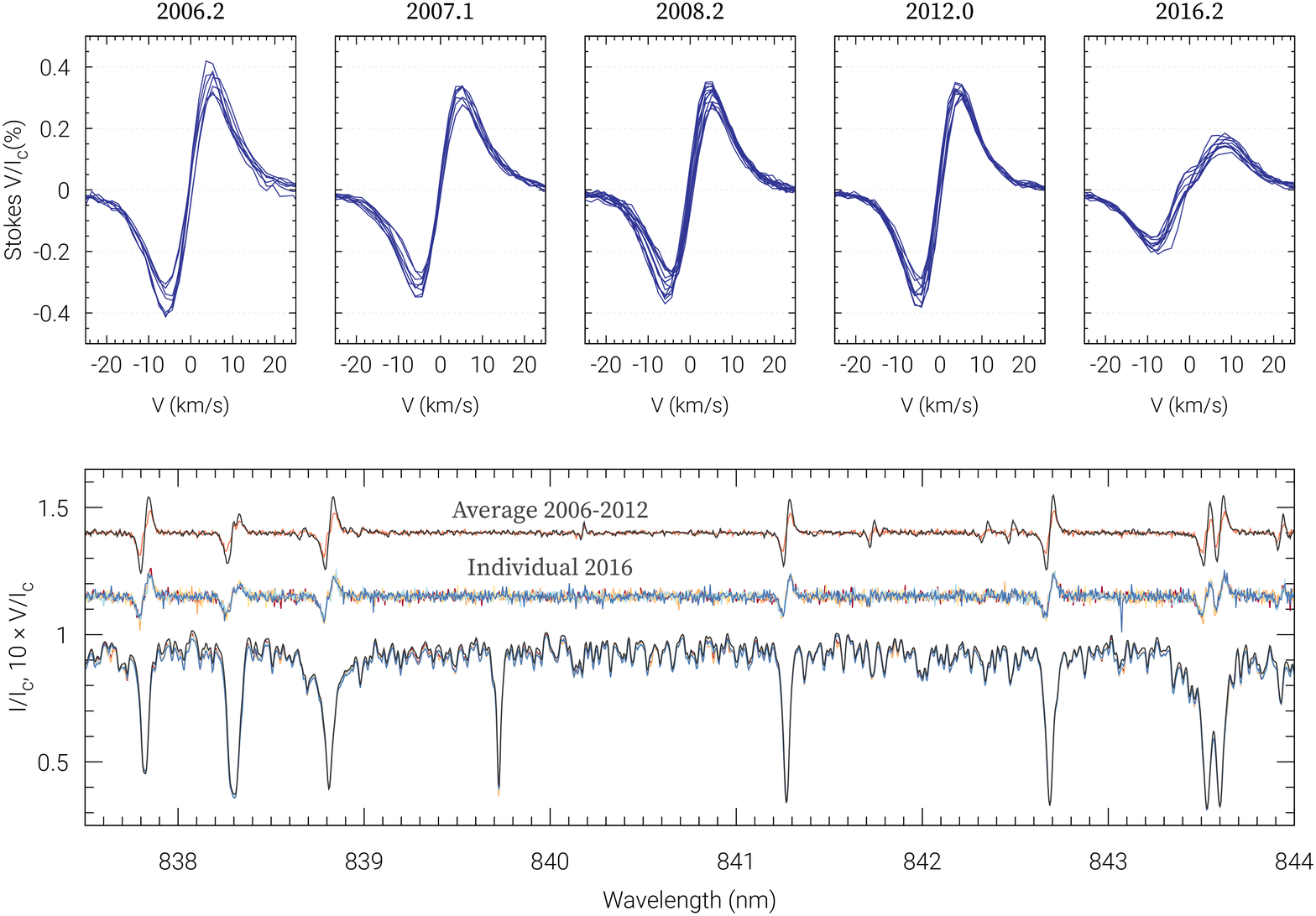}
    \caption{Comparison of archival observations of AD~Leo with our new 2016 observations. The top plot shows the Stokes $V$ LSD profiles for all epochs between 2006 and 2016 (blue solid lines). Epochs are indicated at the top of each subplot. The bottom plot displays: (i) individual Stokes $I$ spectra for our 2016 observations overplotted with the average Stokes $I$ spectrum from the 2006--2012 period at the bottom, (ii) individual Stokes $V$ spectra for our 2016 observations shifted to 1.15, and (iii) the average archival Stokes $V$ spectrum (dark line) compared to the average of our 2016 Stokes $V$ spectra (bright orange line) shifted to 1.4. All Stokes $V$ profiles are magnified by a factor of 10.}
    \label{fig:epochs}
\end{figure*}

Here, we report definite detections for all ten Stokes $V$ LSD profiles, three out of six Stokes $Q$ profiles, and two out of six Stokes $U$ profiles. This is the first ever report of a definite detection of linear Zeeman polarisation signal in spectral lines of an active M dwarf. In addition, one Stokes $Q$ profile and three Stokes $U$ profiles are marginally detected, while there is no detection in the rest of the profiles. Our Stokes $V$ LSD profiles have a median amplitude of $1.7 \times 10^{-3} I_{\rm c}$. The median LSD Stokes $Q$ and $U$ profile amplitudes are around 13 times weaker. In other active cool stars where linear polarisation signal was detected, the Stokes $Q$ and $U$ LSD profiles were found to be 5--10 times weaker \citep{2011ApJ...732L..19K, 2013MNRAS.436L..10R} than the Stokes $V$ profiles. 

\subsection{Secular change of the Stokes $V$ spectrum}
\label{subsection:variability}

We show in Fig.~\ref{fig:epochs} a major change of the circular polarisation spectrum between the archival observations obtained in 2006--2012 and our 2016 observing run. The Stokes $V$ LSD profiles from 2016 have systematically smaller amplitude and broader wings than the corresponding profiles derived from the archival data, as can be seen in the top panel of Fig.~\ref{fig:epochs}. We reiterate that all data have been analyzed using the same procedures. As a consequence, these differences can not be attributed to different details of the analysis. 

To investigate the corresponding change in individual spectral lines, we selected five Stokes $I$ and $V$ spectra from 2016 with the highest signal-to-noise ratios ($343\leq \text{SNR} \leq 410$) and compared them against the mean Stokes $I$ and $V$ spectra obtained by averaging all archival observations with SNR\,$\ge200$. This comparison is shown in the bottom panel of Fig.~\ref{fig:epochs}. Only a small wavelength interval of the ESPaDOnS/Narval spectra is displayed in this figure; the picture is qualitatively the same for all other intervals. It is evident that individual spectral lines exhibit the exact same change as seen for the LSD profiles, which confirms that these changes are intrinsic to the stellar spectrum and not an artefact originating from the LSD procedure. 

It is worthwhile to note that the intensity spectra of AD~Leo do not show any significant changes.

\begin{figure}
	\includegraphics[width=1.0\columnwidth]{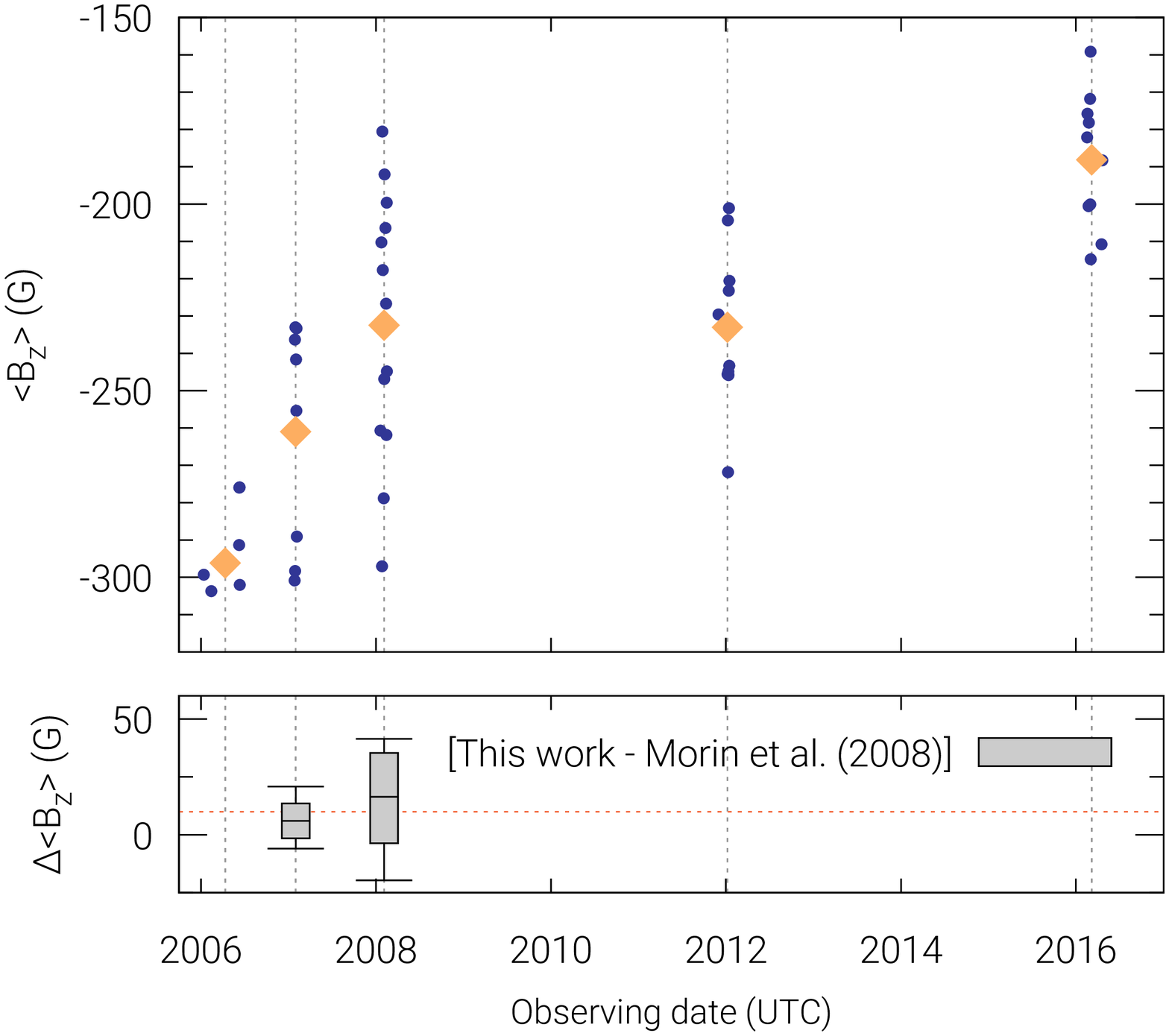}
    \caption{Top panel: mean longitudinal magnetic field {\bz} values (dark blue filled circles) as a function of observing date. The mean date for each observing epoch (2006, 2007, 2008, 2012, and 2016) is indicated with a vertical grey dashed line, and the phase-averaged {\bz} value for each epoch is plotted with large yellow diamonds. Bottom panel: box plots showing the difference of the mean longitudinal magnetic field {\bz} values between this work and \citet{2008MNRAS.390..567M} for the two overlapping epochs (2007 and 2008). The horizontal dashed orange line indicates a difference of 10~G.}
    \label{fig:bz}
\end{figure}

\subsection{Mean longitudinal magnetic field}
\label{subsection:bz}
The mean longitudinal magnetic field {\bz} was computed for each rotational phase from the LSD Stokes $I$ and $V$ profiles according to the formula:
\begin{equation}
\langle B_z \rangle = -\frac{7.145 \times 10^{6}}{\lambda_0 g_{\text{eff}}}\frac{\int v V(v) dv}{\int \left[1 - I(v)\right] dv}
\end{equation}
where {\bz} is measured in G, $v$ is the radial velocity in \kms, $\lambda_0$, in \AA, and $g_\text{eff}$ are, respectively, the mean LSD profile wavelength and mean effective Land\'e factor used for normalisation of the LSD weights \citep{2010A&A...524A...5K}. 

We obtained {\bz} values ranging between $-215$ and $-159$~G with typical uncertainties of 9~G. These measurements are listed for each epoch in column 8 of Table~\ref{table:observation_log}. Additionally, we consistently computed {\bz} values for all archival LSD profiles, using the same velocity range from $-30$ to $+30$ \kms\ for all epochs.

The Stokes $I$ LSD profiles exhibit a slightly different pseudo-continuum level, depending on small variations in normalisation of the spectra, requiring us to scale both the Stokes $I$ and $V$ profiles by a different factor for each year. {\bz} values calculated without rescaling the LSD profiles are on average 130 G higher than {\bz} calculated with the rescaled profiles. 

The evolution of the longitudinal magnetic field of AD~Leo from 2006 to 2016 is presented in Fig.~\ref{fig:bz}. This figure also shows that there is a systematic difference  -- of 10 G on average -- between the {\bz} measurements obtained in our work for 2007--2008 archival spectra and those reported by \citet{2008MNRAS.390..567M}. This discrepancy likely stems from different renormalisation of the Stokes $I$ profiles. A long-term evolution of the mean longitudinal magnetic field can also be distinguished from Fig.~\ref{fig:bz}, showing a steady increase of {\bz} during the 2006--2008 period, followed with comparable values in 2008 and 2012, and a new increase in 2016.

\subsection{Zeeman Doppler imaging}
\label{subsection:zdi}

We carried out a Zeeman Doppler imaging (ZDI) reconstruction of the surface magnetic field of AD~Leo using the Stokes $I$ and $V$ profiles to obtain further insight into the field topology and to compare our 2016 data with the most recent archival dataset from 2012. The two last epochs of the 2016 dataset were discarded as they were acquired more than 20 rotation cycles after the bulk of our observations, which comprise 8 observations spanning 7 rotation cycles. The 2012 map was reconstructed from 9 observations spanning 3.6 rotation cycles.

The ZDI mapping was performed with the {\tt InversLSD} code developed by \citet{2014A&A...565A..83K} and subsequently used for cool active stars by \citet{2016A&A...593A..35R}, \citet{2016A&A...587A..28H} and \citet{2017ApJ...835L...4K}. This code divides the stellar surface into surface elements of roughly equal area, and for each surface element, the local line profiles are computed using the Unno-Rachkovsky analytical solution of the polarised radiative transfer equations \citep{2004ASSL..307.....L}. The average central wavelength and  effective Land\'e factors (670~nm and 1.2, respectively) were adopted according to the LSD line list, and the fiducial line was assumed to split as a Zeeman triplet. The equivalent width of the local profile was adjusted to reproduce the Stokes $I$ LSD profiles. 

%
\begin{figure}
	\includegraphics[width=1.0\columnwidth]{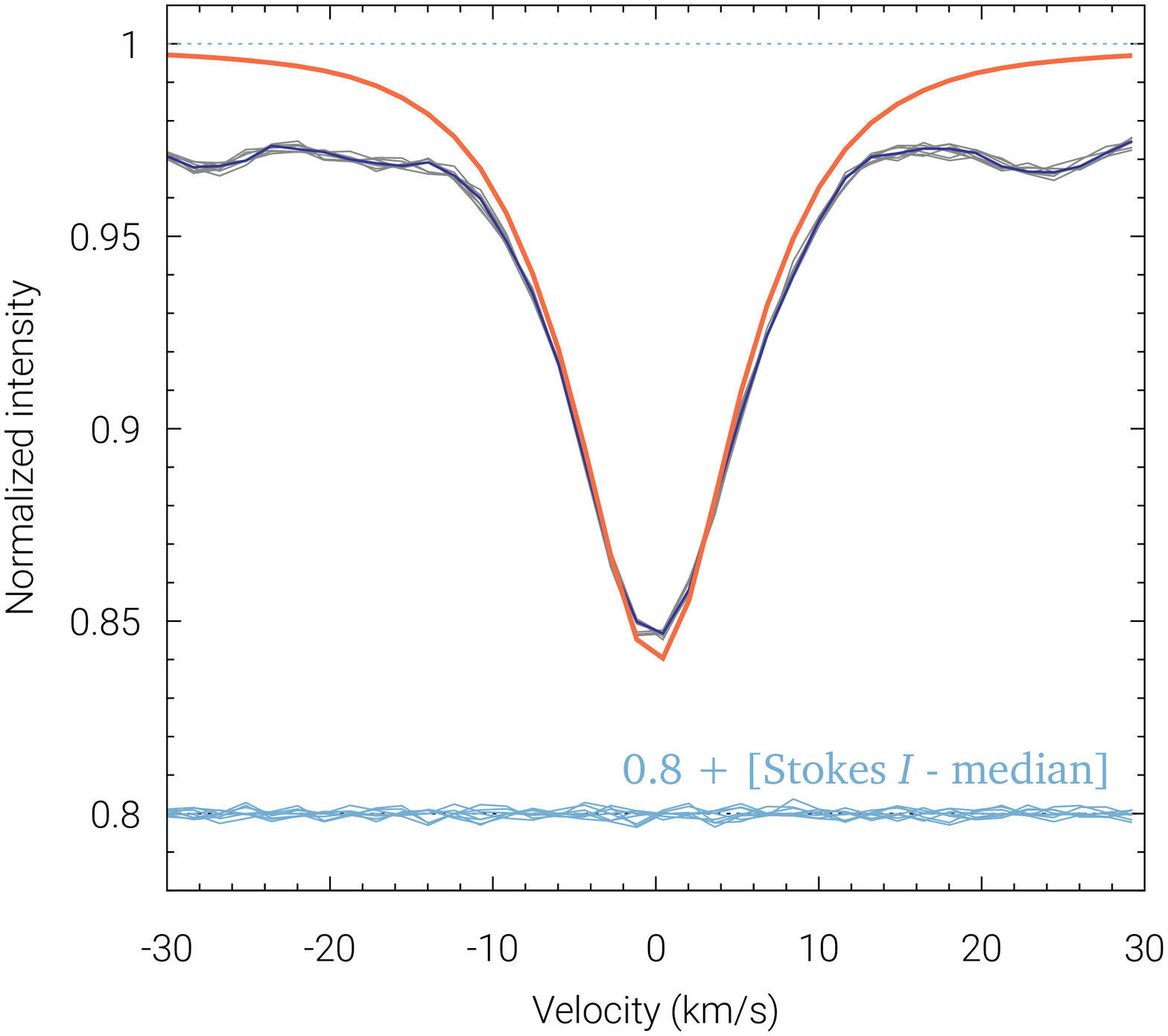}
    \caption{The observed Stokes $I$ LSD profiles for each phase are plotted in light grey, with the median Stokes $I$ LSD profile is overplotted in dark blue. The median Stokes $I$ LSD profile from the ZDI inversion is shown in red. To illustrate the lack of significant rotational variability of the observed profiles, the difference between individual observed Stokes $I$ LSD profiles and the median observed profile is shown in light blue below the line profiles.}
    \label{fig:stokes_I}
\end{figure}

Following \citet{2008MNRAS.390..567M}, we used two additional free parameters, the filling factors $f_I$ and $f_V$, to improve the fit to the observed Stokes $I$ and Stokes $V$ profiles. The physical meaning of these filling factors is that a fraction $f_I$ of each surface element is magnetic, and a fraction $f_V$ of each surface element gives a contribution to the net circular polarisation signature. The definition of the $f_V$ filling factor according to \citet{2008MNRAS.390..567M} is such that the local field strength of the magnetised fraction of a surface element is given by $B/f_V$. The same value of each filling factor is applied to all surface elements across the stellar surface. Following the practice of previous ZDI studies of M dwarfs, we adopted fixed $f_I = 50 \%$ for both studied epochs. This allowed the synthetic LSD Stokes $I$ profiles to match the observed profiles reasonably well. On the other hand, we determined the individual best-fitting $f_V$ filling factors for each epoch individually, by running inversions with a range of different $f_V$ values and selecting the value yielding the minimum $\chi^2$.

Following the practice of previous studies of sharp-line active M dwarfs \citep{2008MNRAS.390..567M,2010MNRAS.407.2269M}, we use only Stokes $V$ LSD profiles in the ZDI analysis and do not attempt to simultaneously fit the details of Stokes $I$ profiles.  The rationale is that it is impossible to simultaneously model satisfactorily the Stokes $I$ and $V$ profiles of M dwarfs using the current framework since no quantitative models incorporating both the global and much stronger local fields have been proposed. Additionally, the Stokes $I$ LSD profiles exhibit a pseudo-continuum level which is several percent below unity, which is attributed to the contribution of molecular lines which are not accounted for in our LSD atomic line mask, meaning that the observed LSD intensity profiles cannot be reproduced as well as circular polarisation profiles. For this reason, we only use the Stokes $I$ LSD profiles to confirm the filling factor $f_I$ and and determine the line parameters used in the Unno-Rachkovsky equations. In addition, in the case of AD Leo, the Stokes $I$ profile shows little rotational variability (see Fig.~\ref{fig:stokes_I}), justifying the assumption of uniform brightness.

\begin{figure*}
	\includegraphics[width=1.9\columnwidth]{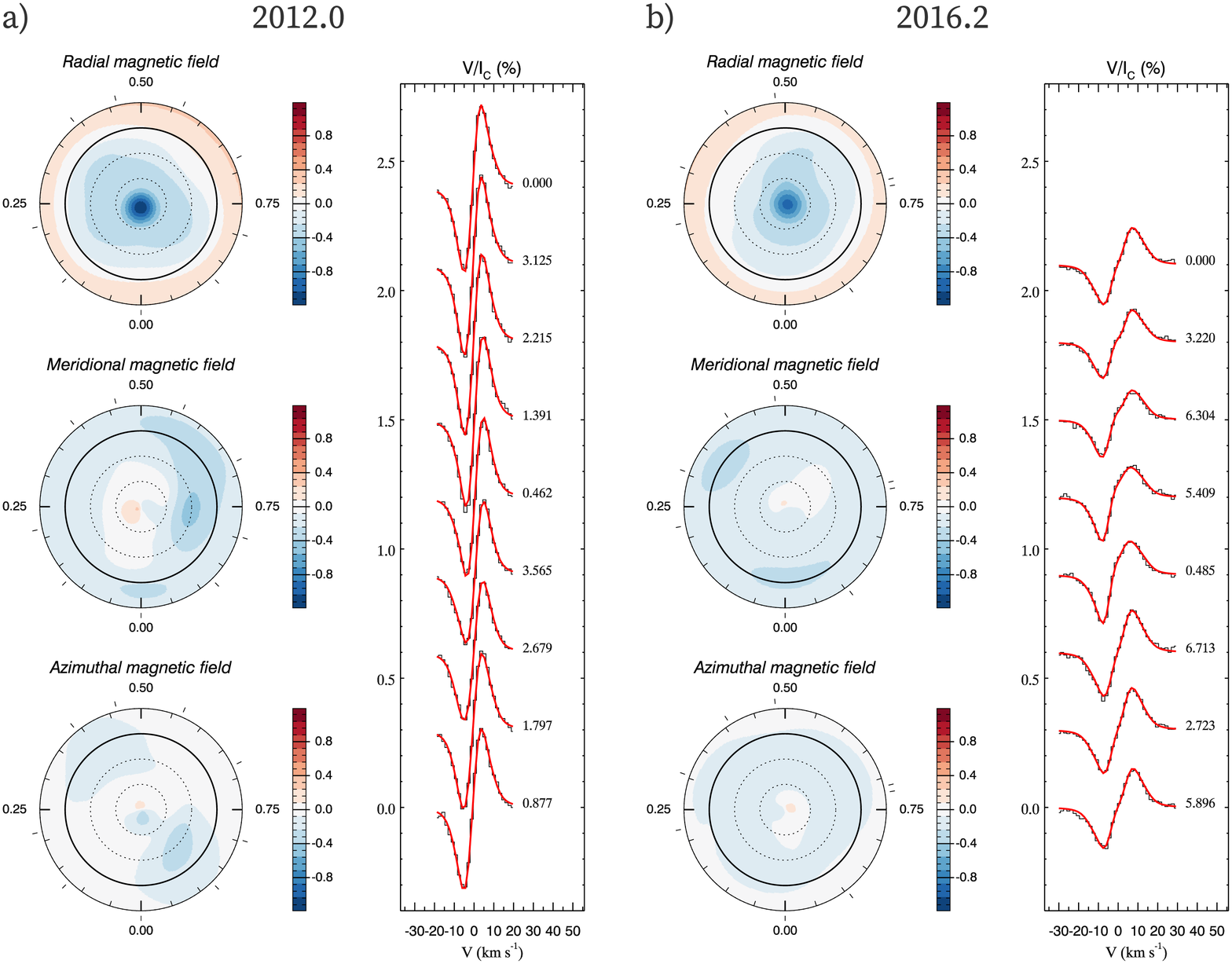}
    \caption{Global surface magnetic field of AD~Leo reconstructed in (a) 2012 and (b) 2016. For each epoch, the flattened polar projections show the radial (top), meridional (middle), and azimuthal (bottom) magnetic field components. In each projection, the stellar surface is shown down to $-30^\circ$ latitude, the thick black circle shows the equator, and the dotted circles correspond to $+30^\circ$ and $+60^\circ$ latitudes.  The colour bar to the right of each map links the colour palette to the magnetic field strength expressed in kG. To the right of the magnetic field maps we show the observed (black histogram) and model (red solid line) LSD Stokes $V$ profiles. These profiles are shifted vertically by a constant offset according to the rotational phase increasing downwards.}
    \label{fig:maps}
\end{figure*}

Since the Stokes $V$ LSD profiles are broader in 2016 than in 2012, we used a wider velocity range for the ZDI modelling of the former profiles, as can be seen in Fig.~\ref{fig:maps}.
%

The magnetic field was represented using a spherical harmonic expansion, with a maximum angular degree $\ell_{\text{max}} = 10$, which we consider to be sufficient given the simple topology of the global magnetic field of AD~Leo. The surface magnetic field inversion was regularised by penalising high $\ell$-modes, essentially favouring simpler field topologies, as described by \citet{2014A&A...565A..83K}. We adopted $v_{\rm e} \sin i = 3$~\kms, an inclination angle $i = 20\degr$, and a rotation period $P_{\text{rot}} = 2.2399$~d as determined by \citet{2008MNRAS.390..567M}. A radial velocity correction of 12.37~\kms, corresponding to the average value determined by \citet{2008MNRAS.390..567M}, was applied to all our spectra.

The resulting surface magnetic field maps and fits to the observed Stokes $V$ LSD profiles are presented in Fig.~\ref{fig:maps}. We observe a quantitative transformation of the surface magnetic field between 2012 and 2016, with a general decrease of the mean magnetic field strength. However, it is interesting to note that a significant reduction of the Stokes $V$ filling factor $f_V$, from 13\% in 2012 to $\le$7\% in 2016, was required in order to fit the observed LSD profiles. This means that while the average strength of the global field component is decreasing, the actual local magnetic field values ($B/f_V$) are higher in our 2016 inversion. This hints that the field component dominating the circular polarisation profiles of AD~Leo became concentrated in smaller areas in 2016 compared to all previous epochs when this star was monitored with high-resolution spectropolarimetry. 

Despite this evolution, the topology of the global magnetic field is mostly unchanged and is still predominantly dipolar (89\% of the total magnetic field energy is in the $\ell=1$ mode in 2012 compared to 94\% in 2016), poloidal (94\% of the field energy is in the poloidal component in 2012 against 91\% in 2016) and axisymmetric (93\% and 97\% of the total energy is in $m<\ell/2$ modes for the 2012 and 2016 magnetic maps, respectively). Table~\ref{table:inversion_parameters} lists the change of the field parameters inferred from the ZDI inversions of the 2012 and 2016 datasets.

Additionally, we compared the observed and synthetic total linear polarisation $P = \sqrt[]{Q^2 + U^2}$ for the first 4 full Stokes epochs. The synthetic $P_{\rm syn}$ was obtained by a forward computation of the Stokes $Q$ and $U$ profiles corresponding to our recovered 2016 ZDI map. The observed $P_{\rm obs}$ was calculated from the observed Stokes $Q$ and $U$ LSD profiles. We find that the ratio $P_{\rm syn} / P_{\rm obs}$ ranges from $2$ to $5.5$, indicating that our ZDI inversion -- run without constraining linear polarisation -- overestimates linear polarisation signatures compared to observations. This behaviour is not unexpected, and would hint at the presence of a magnetic field topology with many small-scale structures. Indeed, it was observed for magnetic Ap stars that the amplitude of linear polarisation profiles decreased if small-scale structures were present on the stellar surface \citep{2004A&A...414..613K,2018A&A...609A..88R}. Interestingly, we note that the stronger Stokes $Q$ and $U$ profiles observed at the rotational cycles 27-28 give rise to an observed total linear polarisation with similar amplitude than $P_{\rm syn}$ computed for the earlier epochs.

\begin{table}
	\centering
	\caption{Magnetic field characteristics obtained from the ZDI inversions of the 2012 and 2016 datasets. }
	\label{table:inversion_parameters}
	\begin{tabular}{rcc} 
		\hline
        Distribution of the		&		&		\\
		magnetic field energy 	& 2012	&2016	\\
		\hline
		$\ell=1$	&89.0 \%	& 94.1\%	\\
        $\ell=2$	& 4.4 \%	& 2.0\%	\\
        $\ell=3$	& 1.3 \%	& 1.1\%	\\
        $\ell=4$ 	& 1.4 \%	& 1.1\%	\\
        $\ell=5$	& 1.3 \%	& 0.7\%	\\
        $\ell=6$	& 0.9 \%	& 0.4\%\\
        $\ell=7$	& 0.7 \%	& 0.2\%\\
        $\ell=8$	& 0.4 \%	& 0.2\%\\
        $\ell=9$	& 0.3 \%	& 0.2\%\\
        $\ell=10$ & 0.2 \% 	& 0.2\%\\
		poloidal	& 94.0\%	& 90.6 \%	\\ 
		axisymmetric ($m<l/2$) 	& 92.6\%	& 97.2\%	\\
        \hline
        Magnetic field characteristics & 2012 & 2016 \\
        \hline
        $f_I$	& 50\%	  & 50\% \\
        $f_V$&  $13^{+1.0}_{-0.5}$\%	& $7^{+0.5}_{-3.0}$\% \\
        average magnetic field	$\overline{B}$	& 0.33 kG	&  0.30 kG \\
        $\overline{B} / f_V$	&2.54  kG	&4.29  kG \\	
        maximum magnetic field $B_{\text{max}}$	& 1.18 kG	&  0.95 kG \\
        $B_{\text{max}} / f_V$		& 9.08 kG	&13.57  kG\\
		\hline
	\end{tabular}
\end{table}
%
\section{Discussion}
\label{section:discussions}
In this article we analysed new spectropolarimetric observations of the active M dwarf star AD Leo. For the first time, circular polarimetry is complemented with linear polarisation observations. We reported definite detections of the linear polarisation signatures in LSD profiles. These are the first detections of a Zeeman linear polarisation signal in the spectral lines of an M dwarf. However, its amplitude turned out to be not as strong relative to the circular polarisation signature as was expected from previous studies of more massive active stars. In general, the linear polarisation proved to be too weak and noisy to be used for constraining the ZDI inversions. 

While AD~Leo is a bright and active M dwarf with some of the strongest circular polarisation signatures, we recorded only relatively weak signals in linear polarisation, even with long exposure times (2400 seconds) at a 4-m class telescope equipped with an excellent optical spectropolarimeter. As a consequence, acquiring spectropolarimetric time series observations of the complete Stokes vector of AD~Leo (and other M dwarfs) suitable for Zeeman Doppler imaging may not be possible with the current generation of instrumentation.

Unexpectedly, we discovered a secular evolution of the amplitude and shape of the circular polarisation profiles -- both in individual spectral lines and in LSD profiles. This change, occurring between 2012 and 2016, indicates a sudden (judging by the stability of the Stokes $V$ signatures during the 2006--2012 period covered by the archival data) transformation of the surface magnetic field of AD Leo. This transformation is reflected in the surface magnetic field maps reconstructed by applying ZDI to the 2012 and 2016 Stokes $V$ profile sets. Magnetic maps show a decrease of the average and maximum magnetic field strength, but do not indicate a significant change of the magnetic field topology, which remains predominantly an axisymmetric dipole. However, the modeling also indicates that magnetic features have become more concentrated between 2012-2016, leading to the observed Stokes $V$ profiles arising from a smaller fraction of the surface.

These results show that long-term spectropolarimetric monitoring of active M dwarfs -- even those with purportedly stable and well-understood global field topologies -- is necessary and can be valuable for bringing observational constraints for dynamo processes in these stars and discriminating between alternative theoretical models. 

Recent direct numerical simulations (DNS) of dynamos in fully convective, stratified spheres have been relatively successful in reproducing some of the observed characteristics of M dwarf magnetic fields. \citet{2013A&A...549L...5G} presented models which developed both dipolar and multipolar magnetic field configuration, with a so-called bistable dynamo regime for fully convective stars, where both simple and complex magnetic fields are found for the same stellar parameters. More recently, \citet{2015ApJ...813L..31Y} presented updated dynamo models that were claimed to reproduce well both the typical dipole-like, global magnetic field topology found with ZDI for many M dwarfs and their much stronger local tangled fields diagnosed by the Zeeman broadening studies. At the same time, they provide no interpretation of the extremely small global field filling factors required by the observational ZDI studies.

The DNS results by \citet{2013A&A...549L...5G} and \citet{2015ApJ...813L..31Y}, however, do not predict the occurrence of magnetic cycles. So, finding a systematic long-term evolution of the dipole-dominated field topology of the kind we report here for AD~Leo represents a puzzle for that theory. On the other hand, based on the mean field dynamo calculations, \citet{2014MNRAS.442L...1K} suggested that the dipolar and multipolar magnetic field configurations would not need to arise from a bistable dynamo, but could instead result from an oscillatory dynamo mechanism, that generates magnetic cycles as in the Sun. Thus, M dwarfs would exhibit observable magnetic cycles -- with periods up to a few decades -- resulting from this oscillatory dynamo. During these magnetic cycles, the star would display either a strong axisymmetric field or a weak non-axisymmetric field, if observed close to the magnetic reversal.

In the case of AD~Leo, we do not observe a fundamental topological transformation of the global field structure. The field seems to become more dipolar and axisymmetric, and weaker on the global scale. But this geometry is, apparently, being produced by increasingly stronger local magnetic features which concentrate in smaller surface areas. According to our knowledge, this behaviour was previously not considered by any global stellar dynamo models.

This is the first spectropolarimetric study of an M dwarf over a period spanning a decade. Previously, \citet{2008MNRAS.390..567M,2010MNRAS.407.2269M} performed spectropolarimetric studies of late- and mid-M dwarfs over a period of a few years. They observed substantial magnetic variability for one of their fully-convective targets, GJ~1245~B, which exhibited a variable multipolar field which changed significantly every year in the 2006--2008 period. Our results for AD~Leo are different in the sense that we have identified a secular change of the field structure in an M dwarf with predominantly dipolar global field topology.

If occurring, magnetic cycles on M dwarfs are expected to have longer periods than in the Sun, perhaps more than a few decades. To observe these cycles, or the lack thereof, and thus refine our understanding of M dwarf magnetism, monitoring of magnetically active M dwarfs over a long period will be essential.

\section*{Acknowledgements}
The authors would like to thank the CFHT Director for awarding Director's Discretionary Time to their project. A.L acknowledges financial support from ESO Director General Discretionary Funding, and would like to thank G. A. J. Hussain and J. Morin for their help and their useful remarks. O.K. acknowledges financial support from the Knut and Alice Wallenberg Foundation, the Swedish Research Council, and the Swedish National Space Board. This research has made use of NASA's Astrophysics Data System Bibliographic Services, and of the SIMBAD database, operated at CDS, Strasbourg, France \citep{2000A&AS..143....9W}.



\bibliographystyle{mnras}
\bibliography{adleo} 


%
%

\bsp	
\label{lastpage}
\end{document}